\documentclass[twocolumn,showpacs,amsmath,amssymb,prd,floatfix]{revtex4}
\usepackage{graphicx}
\usepackage{bm}
\usepackage{amsmath}
\input{colordvi.tex}
\usepackage{color}
\usepackage[varg]{txfonts}

\newcommand{\simgt}{\lower.5ex\hbox{$\; \buildrel > \over \sim \;$}}
\newcommand{\simlt}{\lower.5ex\hbox{$\; \buildrel < \over \sim \;$}}

\begin{document}

\title[Geometrical constraint on curvature with BAO experiments]
{Geometrical constraint on curvature with BAO experiments}

\author{Masahiro Takada$^1$
and
Olivier Dor\'e$^{2,3}$} 
\affiliation{$^1$Kavli Institute for the Physics and
Mathematics of the Universe (Kavli IPMU, WPI),
The University of Tokyo Institutes for Advanced Study, 
The University of Tokyo, Chiba
277-8583, Japan\\
$^2$Caltech M/C 350-17, Pasadena, CA 91125, USA\\
$^3$Jet Propulsion Laboratory, California Institute of Technology, 4800 Oak Grove Drive, Pasadena, California, U.S.A.}

 \begin{abstract}
  The spatial curvature ($K$ or $\Omega_K$) is one of the most
  fundamental parameters of an isotropic and homogeneous universe and has a
  close link to the physics of the  early Universe. Combining the radial and
  angular diameter distances measured via the baryon acoustic
  oscillation (BAO) experiments allows us to 
  unambiguously constrain the curvature.
  The method is primarily based on the metric theory, but is less
  sensitive to
  the theory of structure formation other than the existence of the BAO
  scale and is free of any model of dark energy.
  In this paper, we estimate a best achievable accuracy of constraining
  the curvature with the BAO experiments. We show that an all-sky,
  cosmic-variance-limited galaxy survey covering the Universe up to
  $z\simgt 4$ enables a precise determination of the curvature to an
  accuracy of $\sigma(\Omega_K)\simeq 10^{-3}$. When we assume a model
  of dark energy -- either the cosmological constant or the
  $(w_0,w_a)$ model --  it can achieve a precision of
  $\sigma(\Omega_K)\simeq \mbox{a few}\times 10^{-4}$. These forecasts
  require a high sampling density of galaxies, and are degraded by up to
  a factor of a few for a survey with a finite number density of $\sim
  10^{-3}(h/{\rm Mpc})^3$.
 \end{abstract}

\maketitle

\section{Introduction}

The curvature of the Universe (hereafter denoted as $K$ or $\Omega_K$)
is one of the most fundamental quantities in an isotropic and
homogeneous universe in the context of general relativity (GR)
\citep{WeinbergBook:72}. The curvature also has a close connection to
the physics of the early Universe. An inflationary universe scenario
predicts that the ``apparent'' curvature, which we can infer from an
observable universe, should appear to be close to a flat geometry
($\Omega_K\simeq 0$), even if the exact value is nonzero
\citep{Guth:81,Sato:81}.  If the Universe arose from the decay of a false
vacuum via quantum tunneling, it leads to an open geometry ($K<0$ or
$\Omega_K>0$)
\citep{ColemanDeLuccia:80,Gott:82,Bucheretal:95,Linde:95,Yamamotoetal:95}. In
particular, if the Universe began with ``large-field inflation''
\citep{Lyth:97} -- which predicts that the primordial gravitational wave
is as large as can be observed in the cosmic microwave background (CMB)
anisotropies -- or if the universe began with ``just enough inflation'' in
a landscape or multiverse picture, the curvature can be large enough to be
measurable $(\Omega_K\sim 10^{-4}$--$10^{-2})$
\citep{Weinberg:89,GuthNomura:12,KlebanSchillo:12,Freivogeletal:14,Boussoetal:14}.
Further, an addition of the super-curvature perturbation in an
open-inflation scenario might resolve the large-scale CMB anomalies
\citep{LiddleCortes:13,Kannoetal:13,Whiteetal:14}. On the other hand,
if a closed curvature ($K>0$ or $\Omega_K<0$) is found, it gives rise to
a challenge for the inflationary scenario: the Universe needs to emerge from
a specific initial condition of closed curvature
\citep{HartleHawking:83,Linde:95}. Thus it is important to constrain the
curvature from cosmological observations in order to obtain a clue to
the physics of the early universe 
\citep[also see Ref.][for a thorough review of
various cosmological probes]{Weinbergetal:12}.

Observations of the CMB have led to the precise measurement of the angular
diameter distance to the last scattering surface, preferring a flat
geometry as expected in an inflationary scenario
\citep[e.g.,][]{WMAP7,Planck2015:cosmo}. However, the constraint rests
on the use of the standard cosmological model such as the cosmological
constant, cold-dark-matter dominated model ($\Lambda$CDM model), based
on GR and the nearly adiabatic initial conditions. If the assumptions
are relaxed -- for instance, if a generalized model of dark energy is
employed -- the CMB constraint on the curvature is largely degraded
\citep{Ichikawaetal:06}. The baryonic acoustic oscillation (BAO)
provides us with an alternative, powerful geometrical probe, allowing
one to constrain the cosmological distances via measurements of the
galaxy clustering pattern in redshift and angular directions,
respectively
\citep{SeoEisenstein:03,HuHaiman:03,Eisensteinetal:05,Percivaletal:07}. The
BAO experiments are shown to be robust against various astrophysical
systematic effects such as the galaxy bias uncertainty
\citep{Eisensteinetal:07,SeoEisenstein:07}. The current state-of-the-art
measurements were done using the Sloan Digital Sky Survey III Baryon
Oscillation Spectroscopic Survey
\citep{BOSSBAO:12,Andersonetal:14,BOSSBAO:14}, achieving a percent
precision of the distance measurement at $z\simeq 0.57$. Various
wide-area spectroscopic galaxy surveys are planned that aim to achieve
precise BAO measurements up to higher redshifts: the Subaru Prime Focus
Spectrograph (PFS) \citep[][]{Takadaetal:14}, the Dark Energy
Spectrograph Instrument (DESI) \citep[][]{DESI}, the ESA Euclid
satellite mission \footnote{http://sci.esa.int/euclid/}, and the NASA
WFIRST mission \citep{WFIRST:15}.

The BAO method is unique in that it can constrain the radial (more
exactly the Hubble expansion rate) distance as well as the angular
diameter distance at the redshift of the galaxy survey, while other
geometrical probes such as supernovae and gravitational lensing can
probe the luminosity or angular distances (and not the radial distance). The
relation between the radial and angular diameter distances is purely
geometrical and specified by the curvature; if the Universe has a
nonzero curvature, the two distances differ.  The relation holds for
any theory of gravity or dark energy, and rests on the metric theory of
a homogeneous and isotropic space, which is a maximally symmetric
spacetime described by the Friedmann-Robertson-Walker (FRW) metric
\citep{WeinbergBook:72}.  Hence the purpose of this paper is to estimate
the fundamental accuracy of estimating the curvature parameter with the
BAO experiments \citep[see Refs.][for the previous works based
on the similar motivation]{Knox:06,Bernstein:06}.
To do this, we will assume a cosmic-variance-limited
galaxy survey, namely full-sky coverage and a sufficiently high number
of density of sampled galaxies.

The structure of this paper is as follows. In Sec.~\ref{sec:method} we
discuss the methodology based on the Fisher matrix information
formalism, and briefly review the BAO method. In Sec.~\ref{sec:results}
we show the main results. Section~\ref{sec:discussion} is devoted to
discussion. Unless stated otherwise, we will adopt as fiducial model a
flat $\Lambda$CDM cosmology with $\Omega_{\rm m0}=0.27$,
$\Omega_{\Lambda}=0.73$ and the Hubble parameter $h=0.71$.

\section{Geometrical estimation of the curvature with BAO distances}
\label{sec:method}

\subsection{Cosmological distances}

We assume that the Universe is 
statistically isotopic and homogeneous. The spacetime structure of such
an universe is described by the FRW metric \citep{WeinbergBook:72}.
With the metric theory, solving the light propagation in an expanding
universe yields a relation between cosmological distances and
redshift. The comoving radial distance is given in terms of the
integral of the Hubble expansion rate:
\begin{equation}
 D_C(z)\equiv \int_0^z\!\frac{dz'}{H(z')},
  \label{eq:dc}
\end{equation}
where the Hubble expansion rate is given by the time derivative of the scale
factor as $H\equiv d{\ln a}/dt$, and $z$ is the redshift, given as
$1+z\equiv 1/a$ where we employed the convention $a(t_0)=1$ today.
The metric theory also gives a geometrical relation between the radial
and angular diameter distances:
\begin{equation}
D_A(z)=\left\{
\begin{array}{ll}
{\displaystyle \frac{1}{\sqrt{-K}}\sinh\sqrt{-K}D_C(z)} & (K<0),\\
D_C(z) & (K=0),\\
{\displaystyle \frac{1}{\sqrt{K}}\sin\sqrt{K}D_C(z)} & (K>0)\\
\end{array}
       \right.
\label{eq:da_dc}
\end{equation}
The different equations are for different geometries of the Universe;
open, flat and close geometries, respectively. The curvature $K$ is in
units of $1/{\rm distance}^2$. The CMB experiments
\citep[e.g., Ref.][]{WMAP7} give stringent constraints on the curvature,
implying $|\Omega_K|\equiv |K|/H_0^2\ll 1$, where $\Omega_K$ is the
curvature parameter and $1/H_0$ is the Hubble radius today.
Note that throughout this paper we use the {\em comoving} angular
diameter distance, rather than the physical distance, and the two are
related as $D_A^{\rm phys}(z)=D_A(z)/(1+z)$.
For redshifts
relevant for galaxy surveys,
where $\sqrt{|K|}D_C(z)\ll 1$ holds,
Eq.~(\ref{eq:da_dc}) can be expanded around
$\sqrt{|K|}D_C(z)=0$, yielding the approximation
\begin{equation}
D_A(z)\simeq D_C(z)
\left[1-\frac{1}{6}K D_C(z)^2
\right].
\label{eq:da-dc}
\end{equation}
Both $D_A(z)$ and $D_C(z)$ are observables of the BAO experiments for
each redshift slice. In the following we will use the above equation to
estimate the accuracy of estimating the curvature from BAO information
without making many assumption about the theory of structure formation
(other than the existence of the BAO scale in the distribution of galaxies)
or any model of dark energy.

\subsection{Estimator of the curvature and its covariance}

Suppose we have the BAO distance measurements in $N_s$ redshift bins,
without any gap, over a range of redshift from today up to a maximum
redshift $z_{\rm max}$: $\hat{D}_A(z_i)$ and $\hat{D}_H(z_i)$ for
$i=1,2, \dots, N_s$ ($z_1=0$ and $z_{N_s}=z_{\rm max}$). Here $D_H(z)$
is the comoving Hubble distance: $D_H(z)\equiv 1/H(z)$.

The radial distance at the $i$th redshift bin $z_i$ can be estimated by
combining the measured Hubble distances over a redshift range
$z=[0,z_i]$ (see Eq.~\ref{eq:dc}):
\begin{equation}
 \hat{D}_{C,i}\simeq \sum_{z_j<z_i} \hat{D}_{H,j}\Delta z_j,
\end{equation}
where we have introduced the notations, $\hat{D}_{C,i}\equiv
\hat{D}_C(z_i)$ and $\hat{D}_{H,i}\equiv \hat{D}_H(z_i)$.
In practice one might want to use a more sophisticated method to
estimate $\hat{D}_C(z)$ to avoid inaccuracy due to the discrete
summation, e.g. by assuming that $H(z)$ is modeled by a polynomial
function of redshift and then estimating the coefficients from fitting
of the function
to the measured $H(z)$.  Here we assume a discrete summation
for simplicity.

From Eq.~(\ref{eq:da-dc}), an estimator of the curvature parameter from measurements of the radial
and angular diameter distances at the $i$th redshift bin can be given
as
\begin{equation}
\hat{K}_i \equiv 6\frac{
\hat{D}_{A,i}-\hat{D}_{C,i}}{\hat{D}_{C,i}^3}.
\end{equation}
Combining the measurements at different redshift bins, we can define the
chi-square ($\chi^2$) to estimate the curvature parameter as
\begin{equation}
\chi^2\equiv \sum_{i,j=1}^{N_s}(\hat{K}_i-K)
\left[\bm{C}^{KK}\right]^{-1}_{ij}(\hat{K}_j-K),
\label{eq:chi2}
\end{equation}
where $K$ is the underlying true curvature, treated as a free model
parameter in the $\chi^2$ fitting.  $[\bm{C}^{KK}]^{-1}$ denotes the
inverse of the covariance matrix $\bm{C}^{KK}$ defined as
\begin{eqnarray}
\frac{\bm{C}^{KK}_{ij}}{36}
&=&\frac{1}{D_{C,i}^3D_{C,j}^3}\bm{C}^{D_AD_A}_{ij}
-\frac{1}{D^3_{C,i}}\left[3\frac{D_{A,j}}{D_{C,j}^4}
-2\frac{1}{D_{C,j}^3}\right]\bm{C}^{D_AD_C}_{ij}\nonumber\\
&&-\frac{1}{D_{C,j}^3}\left[3\frac{D_{A,i}}{D_{C,i}^4}
-2\frac{1}{D_{C,i}^3}\right]\bm{C}^{D_A D_C}_{ji}
\nonumber\\
&&+\left[3\frac{D_{A,i}}{D_{C,i}^4}
-2\frac{1}{D_{C,i}^3}\right]
\left[3\frac{D_{A,j}}{D_{C,j}^4}
-2\frac{1}{D_{C,j}^3}\right]
\bm{C}^{D_CD_C}_{ij}.
\label{eq:covKK}
\end{eqnarray}
Here the covariance matrices such as $\bm{C}^{D_CD_C}$ describe
statistical uncertainties of the BAO observables -- the Hubble and
angular-diameter distances -- given BAO measurements of
a galaxy survey:
\begin{eqnarray}
\bm{C}^{D_CD_C}_{ij}&\equiv& {\rm Cov}[\hat{D}_{C,i},\hat{D}_{C,j}]
 =\sum_{m<i}\sum_{n<j}\bm{C}^{D_HD_H}_{mn}\Delta z_m\Delta z_n,
\nonumber\\
 \bm{C}^{D_CD_A}_{ij}
  &\equiv & {\rm Cov}[\hat{D}_{C,i},\hat{D}_{C,j}]
  =\sum_{m<i} \bm{C}^{D_HD_A}_{mj}\Delta z_m,
\end{eqnarray}
and so on.  Around a flat-geometry universe, where $D_A\simeq D_C$, the
covariance matrix is approximated as
\begin{eqnarray}
\frac{ D_{A,i}^3D_{A,j}^3}{36}\bm{C}^{KK}_{ij}&\simeq& 
\bm{C}^{D_AD_A}_{ij}-\left[
\bm{C}^{D_A D_C}_{ij}+\bm{C}^{D_A D_C}_{ji}
\right]+\bm{C}^{D_C D_C}_{ij}.
\end{eqnarray}

The statistical uncertainty of the curvature estimation is given from the
second derivatives of $\chi^2(K)$ (Eq.~\ref{eq:chi2}) with respect to the
true curvature:
\begin{equation}
\sigma^2(K)=\frac{1}{\sum_{i,j=1}^{N_s}\left[\bm{C}^{KK}\right]^{-1}_{ij}},
\label{eq:sigmaK}
\end{equation}
where $[\bm{C}^{KK}]^{-1}$ is the inverse of Eq.~(\ref{eq:covKK}).  We
will use the above equation to estimate the accuracy of the curvature
estimation for a given galaxy survey.

\subsection{BAO}
\label{sec:BAO}

In this subsection, assuming a hypothetical galaxy survey, we derive the
Fisher information matrix of the Hubble and angular diameter distances
from the BAO measurements.

The two-point correlation function or the Fourier-transformed
counterpart -- the power spectrum -- is 
measured as a function of
the separation lengths between paired galaxies. In this procedure, the
position of each galaxy needs to be inferred from the measured redshift
and angular position.  Then the separation lengths perpendicular and
parallel to the line-of-sight direction from the measured quantities are
given as $r_\perp\propto \Delta \theta$ and $r_\parallel\propto \Delta
z$, where $\Delta \theta $ and $ \Delta z$ are the differences between
the angular positions and the redshifts of the paired galaxies. For this
conversion, we need to assume a reference cosmological model to relate
the observables ($\Delta \theta$, $\Delta z$) to the quantities
$(r_\perp,r_\parallel)$. Thus, the wave numbers are given as
\begin{equation}
k_{\perp,{\rm ref}}= \frac{D_{A}(z)}{D_{A, {\rm
 ref}}(z)}
k_{\perp},
\hspace{1em}
k_{\parallel, {\rm ref}}=\frac{D_H(z)}{D_{H,{\rm ref}}(z)}
k_\parallel.
\end{equation}
The quantities with the subscript ``ref'' are the quantities estimated from
the observables assuming a ``reference'' cosmological model, and the
quantities without the subscript are the underlying true values. Since
the reference cosmological model assumed generally differs from the
underlying true cosmology, it causes an apparent distortion in the
two-dimensional pattern of galaxy clustering. In principle, this could
be measured using only the isotropy of clustering statistics -- the
so-called Alcock-Paczynski (AP) test
\cite{AP:79} -- but a more robust measurement of both $D_A(z)$ and
$D_H(z)$ can be obtained by searching for the ``common'' BAO scales in
the pattern of galaxy clustering, as the standard ruler, in combination
with the CMB constraints
\citep{SeoEisenstein:03,HuHaiman:03,EisensteinWhite:04}.

We will use the currently standard $\Lambda$CDM model as a guidance for
the parameter dependence of our constraints and as an effective
realistic description of the galaxy clustering. Nevertheless, we would
again like to emphasize that the methodology proposed in this paper
relies only on an FRW metric theory and the existence of the BAO standard
ruler. To be more quantitative, the redshift-space galaxy power spectrum
at redshift $z$
is given in the linear regime as
\begin{eqnarray}
 P_{g,s}(k_{\perp,{\rm ref}},k_{\parallel,{\rm ref}}; z)&&\nonumber\\
&&\hspace{-8em}=\frac{D_{A,{\rm
 ref} }(z)^2 D_{H, {\rm ref}(z)}}
{D_A(z)^2D_H(z)}
\left[1+\beta(z)\frac{k_{\parallel}^2}{k^2}\right]^2b_g^2P^L_m(k;z)
  +P_{\rm
sn},
\label{eq:Pg}
\end{eqnarray}
where $b_g$ is the linear bias parameter, $\beta$ is the linear
redshift-space distortion parameter, defined as $\beta\equiv
(1/b_g)\left.d\ln D/d\ln a\right|_z$ \cite{Kaiser:87}, $D$ is the linear
growth rate, $P^L_m(k)$ is the linear mass power spectrum, and $P_{\rm
sn}$ is a constant number and is a parameter to model the residual shot
noise.  Throughout this paper we assume $b(z_i)=1$ as the fiducial value
in each redshift bin for simplicity. This is a conservative assumption,
because most galaxies at higher redshift are very likely to be biased
tracers with $b>1$. When deriving the BAO geometrical constraints we
will marginalize over the effect of galaxy bias uncertainty.

To make the parameter forecast,
we employ the method developed in
Ref.~\cite{SeoEisenstein:07}. In this method, we include
the smearing effect of the BAO features due to the bulk flow of galaxies
in large-scale structure
\cite{Matsubara:08,Taruyaetal:09,NishimichiTaruya:11}. For the BAO
survey of multiple redshift bins, the Fisher information matrix of model
parameters can be computed as
\begin{eqnarray}
&&\hspace{-4em}F^{\rm galaxy}_{\alpha\beta}=\sum_{z_i}\int_{-1}^1\!d\mu\int^{k_{\rm max}}_{k_{\rm
 min}}\!\frac{2\pi k^2dk}{2(2\pi)^3}
 \nonumber\\
 &&\hspace{-1em}\times
\frac{\partial \ln P_{g, s}(k,\mu; z_i)}{\partial
 p_\alpha}
\frac{\partial \ln P_{g ,s}(k,\mu; z_i)}{\partial
 p_\beta}\nonumber\\
&&\hspace{-1em}
\times V_{\rm eff}(k; z_i)
\exp\left[-k^2\Sigma_\perp^2 -k^2\mu^2(\Sigma_\parallel^2-\Sigma_\perp^2)\right],
\label{eq:fisher}
\end{eqnarray}
where $\mu$ is the cosine between the wave vector and the line-of-sight
direction, $\mu\equiv k_\parallel/k$; $\sum_{z_i}$ is the sum over
different redshift bins; $\partial P_{g, s}/\partial p_\alpha$ is the
partial derivative of the galaxy power spectrum (Eq.~\ref{eq:Pg}) with
respect to the $\alpha$th parameter around the fiducial cosmological
model; and the effective survey volume $V_{\rm eff}$ and the Lagrangian
displacement fields $\Sigma_\parallel$ and $\Sigma$ to model the
smearing effect are given
as
\begin{eqnarray}
V_{\rm eff}(k,\mu;z_i)&\equiv&
 \left[\frac{\bar{n}_g(z_i)P_{g, s}(k,\mu;z_i)}
{\bar{n}_g(z_i)P_{g, s}(k,\mu; z_i)+1}\right]^2V_{\rm
 survey}(z_i),\\
\Sigma_{\perp}(z)&\equiv & c_{\rm rec}D(z)\Sigma_0, \\
\Sigma_{\parallel}(z)&\equiv & c_{\rm rec}D(z)(1+f_g)\Sigma_0.
\label{eq:sigma}
\end{eqnarray}
Here $V_{\rm survey}(z_i)$ is the comoving volume of the redshift slice
centered at $z_i$; the present-day Lagrangian displacement field is
$\Sigma_0=11~h^{-1}{\rm Mpc}$ for $\sigma_8=0.8$ \cite{Eisensteinetal:07}; 
$D(z)$ is the growth
rate normalized as $D(z=0)=1$; $f_g=d\ln D/d\ln a $. The parameter
$c_{\rm rec}$ is a parameter to model the reconstruction method of the
BAO peaks (see below). 
In
Eq.~(\ref{eq:fisher}), we take the exponential factor of the smearing
effect outside of the derivatives of $P_{\rm g,s}$. This is equivalent
to marginalizing over uncertainties in $\Sigma_\parallel$ and
$\Sigma_\perp$.
The growth rate in $\Sigma_{\parallel}$ or $\Sigma_\perp$ takes into
account the smaller smearing effect at higher redshift due to the reduced
evolution of large-scale structure.  For the parameters, we include the
cosmological parameters, the distances in each redshift slice, and the
nuisance parameters:
\begin{eqnarray}
p_{\alpha}&=&\{\Omega_{\rm m0}, A_s, n_s, \alpha_s, \Omega_{\rm m0}h^2,
\Omega_{\rm b0}h^2, D_A(z_i), 
\nonumber\\
&& \hspace{6em}
D_H(z_i), b_g(z_i), 
\beta(z_i), P_{\rm
sn}(z_i)  \}, 
\label{eq:parameters}
\end{eqnarray}
where $A_s$, $n_s$ and $\alpha_s$ are parameters of the primordial power
spectrum, $A_s$ is the amplitude of the primordial curvature
perturbation, and $n_s$ and $\alpha_s$ are the spectral tilt and the
running spectral index.  The set of cosmological parameters determines
the shape of the linear power spectrum.
For the
$k$ integration, we set $k_{\rm min}=10^{-4}~h/{\rm Mpc}$ and $k_{\rm
max}=0.5~h/{\rm Mpc}$ for all the redshift slices, but the exponential factor in
Eq.~(\ref{eq:fisher}) suppresses the information from the nonlinear
scales. The Fisher parameter forecasts depend on the fiducial
cosmological model for which we assumed that the model is consistent with the
WMAP 7-year data \cite{WMAP7}. For a galaxy survey of $N_s$ redshift
bins, the number of model parameters is in total $6+5\times N_s$. 

Furthermore, we assume the BAO reconstruction method in
Ref.~\cite{Eisensteinetal:07}.  Since the peculiar velocity field of
galaxies in large-scale structure can be inferred from the measured
galaxy distribution, the inferred velocity field allows for pulling back
each galaxy to its position at an earlier epoch and then reconstructing
the galaxy distribution more in the linear regime. As a result, one can
correct to some extent the smearing effect in Eq.~(\ref{eq:fisher}) and
sharpen the BAO peaks in the galaxy power spectrum. Padmanabhan et
al. \citep{Padmanabhanetal:12} implemented this method to the real data,
SDSS DR7 LRG catalog, and showed that the reconstruction method can
improve the distance error by a factor of 2. The improvement was
equivalent to reducing the nonlinear smoothing scale from $8.1$ to
$\Sigma_{\rm nl}=4.4~h^{-1}{\rm Mpc}$, about a factor of 2 reduction in the
displacement field. To implement this reconstruction method requires a
sufficiently high number density of the sampled galaxies in order to
reliably infer the peculiar velocity field from the measured galaxy
distribution \citep[see Ref.][for the latest result]{Andersonetal:14}.  In
the Fisher matrix calculation, we used $c_{\rm rec}=0.1$ for an
implementation of the reconstruction method \footnote{We confirmed that 
if we assumed $c_{\rm rec}=0.5$ we can roughly reproduce the distance
measurement accuracy for the SDSS LRGs as found in Ref.~\cite{Padmanabhanetal:12}}.

The BAO reconstruction has so far been successful in substantially
improving the accuracies of both the angular diameter distance and
radial Hubble distance estimations \citep[see Ref.][for a recent
review]{Seoetal:15}. The systematic errors can arise from an imperfect
treatment of nonlinear effects such as nonlinearities in the density
field and the nonlinear motions of galaxies inside massive halos. A
further refinement of the reconstruction method will be required, by
using mock catalogs of galaxy surveys, in order to achieve the ultimate
precision \citep[e.g.,][for such an attempt]{Seoetal:15}. A striking
advantage of using a higher redshift galaxy survey is that the nonlinear
effects are weaker at higher redshifts where large-scale structure 
evolves less, and it will therefore allow a more robust reconstruction
to obtain unbiased measurements of the BAO distances.

In the following forecast, we assume the BAO experiments combined with
the CMB constraints expected from the Planck satellite:
\begin{equation}
\bm{F}=\bm{F}^{\rm CMB} + \bm{F}^{\rm galaxy},
\end{equation}
where $\bm{F}_{\rm CMB}$ is the Fisher matrix for the CMB measurements.
We employ the method in Ref.~\cite{Takadaetal:14} to compute the CMB
Fisher matrix, where we assume the standard $\Lambda$CDM model for the
physics prior to recombination that determines the sound horizon scale
or the BAO scale.

To estimate the accuracy of the distance estimation with the BAO
experiments, we first invert the Fisher matrix $\bm{F}$ and then use the
submatrix including elements of the Hubble and angular diameter
distances:
\begin{eqnarray}
&&   \bm{F} \overset{\mbox{invert}}{\Longrightarrow}
 [\bm{F}]^{-1}
 \overset{\mbox{use submatrix}}{\Longrightarrow}
  [\bm{F}]^{-1}_{MN}
\end{eqnarray}
where the indices $M, N$ denote the elements including the Hubble and
angular diameter distances.  The submatrix $F_{MN}$
gives the error covariance
matrices of distances, $D_A(z_i)$ and $D_H(z_i)$, after marginalizing
over other parameters (6 cosmological parameters and nuisance
parameters). Hence the dimension of $F_{MN}$ is $(2N_s)\times(2N_s)$.
The covariance matrices of $D_A(z_i)$ and $D_H(z_i)$ and the
cross-covariance matrix are
\begin{eqnarray}
&&\bm{C}^{D_AD_A}_{ij} \equiv
[\bm{F}^{-1}]_{D_A(z_i)D_A(z_j)},\nonumber\\
&&\bm{C}^{D_AD_H}_{ij} \equiv
[\bm{F}^{-1}]_{D_A(z_i)D_H(z_j)},\nonumber\\
&&\bm{C}^{D_HD_H}_{ij} \equiv
 [\bm{F}^{-1}]_{D_H(z_i)D_H(z_j)}.
 \label{eq:covDD}
\end{eqnarray}
We use these covariances in Eq.~(\ref{eq:covKK}) to estimate the accuracy
of the curvature estimation.

\section{Results}
\label{sec:results}

\begin{figure*}
    \centering
    \includegraphics[width=5in]{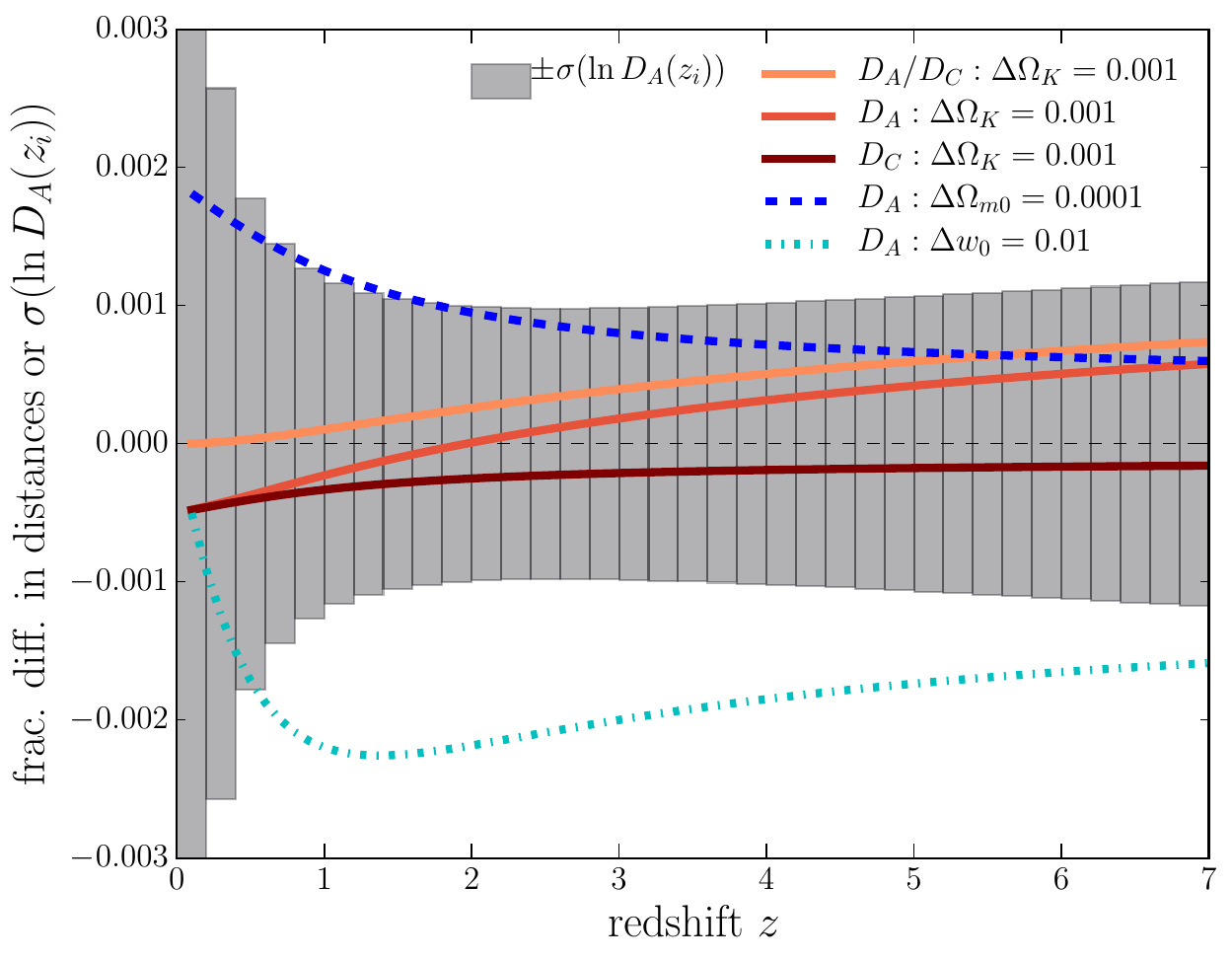}
 \caption{The shaded region in each redshift bin, with width $\Delta
 z=0.2$, denotes the fractional error of the angular diameter distance
 determination $D_A(z_i)$, expected for a cosmic-variance-limited galaxy
 survey with full-sky coverage ($f_{\rm sky}=1$) and a sufficiently high
 number density of galaxies in each redshift over the redshift range
 from $z=0$ to $z_{\rm max}=7$. The errors include marginalization over
 other parameters (see text for details).  For comparison, the solid
 curves show the fractional difference in the radial or angular diameter
 distances, $D_C(z)$ or $D_A(z)$, for an open-geometry universe with
 $\Omega_K=0.001$ compared to the fiducial flat universe, with the dark
 matter density and the cosmological constant being fixed. The curvature parameter
 alters $D_C(z)$ and $D_A$ in different ways and the ratio, $D_A/D_C$,
 displays a characteristic redshift dependence. To be more precise, the
 fractional difference is defined, e.g. as
 $D_A(z;\Omega_K=0.001)/D_{A,{\rm fid}}(z; \Omega_K=0)-1$.  The dashed
 and dot-dashed curves show the fractional difference when changing
 $\Omega_{\rm m0}$ and the dark energy equation of state $w_0$, by
 $\Delta\Omega_{\rm m0}=0.0001$ and $\Delta w_0=0.01$, within a flat
 universe. In this case, the changes keep $D_A(z)=D_C(z)$.}
 \label{fig:dist}
\end{figure*}
\begin{figure*}
    \centering
    \includegraphics[width=5in]{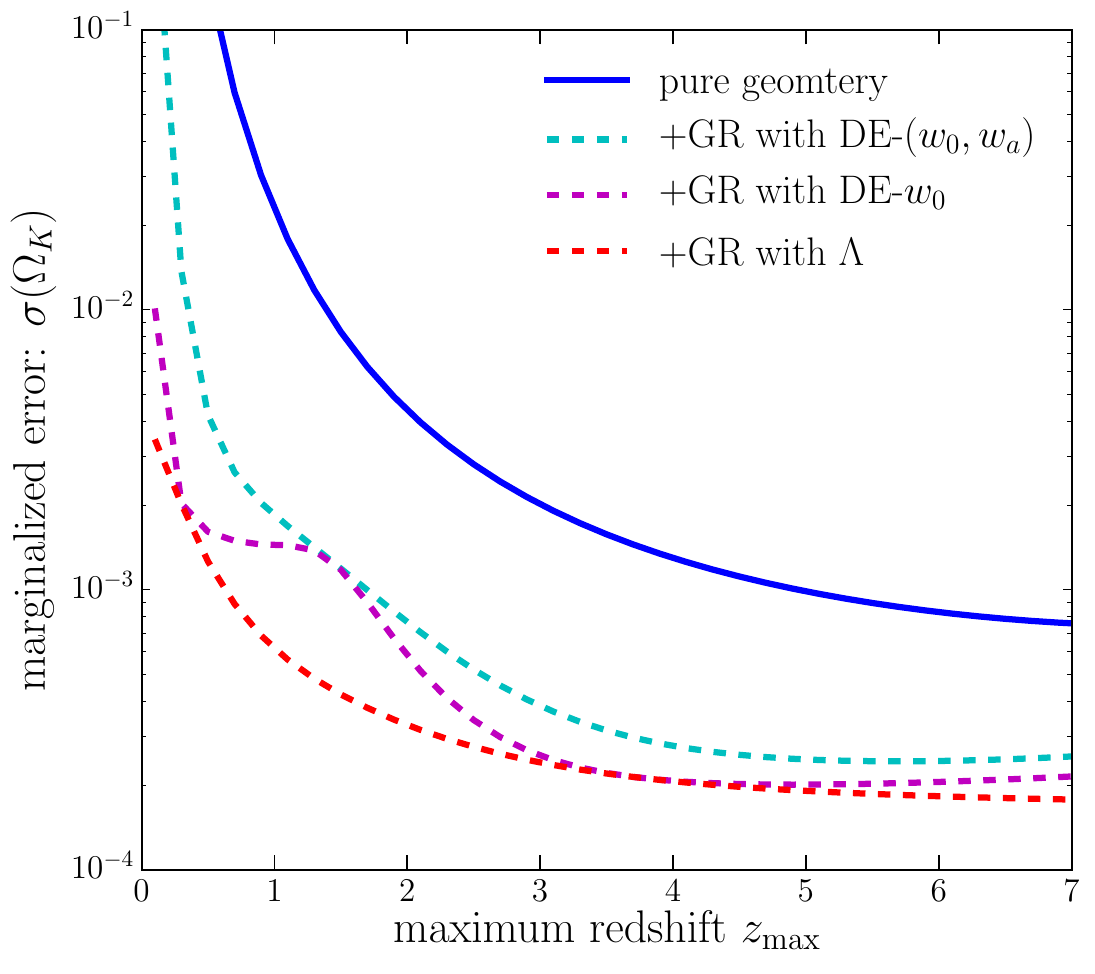}
 \caption{An expected accuracy of the curvature parameter determination
 with the BAO measurements for a cosmic-variance-limited galaxy survey
 as in Fig.~\ref{fig:dist}. Here we consider the galaxy survey
 covering the redshift range $0\le z\le z_{\rm max}$. The top curve
 shows the purely geometrical constraint on $\Omega_K$
 (Eq.~\ref{eq:covKK}), giving the fundamental limit in a sense that the
 constraint is free of any gravity theory or dark energy model and rests
 on the geometrical nature of the four-dimensional spacetime of the
 expanding homogeneous and isotropic universe.  The
 cosmic-variance-limited
 survey can achieve a precision better than $0.1\%$ such as
 $\sigma(\Omega_K)\simeq 0.0008$. If we assume Einstein gravity (GR)
 and employ a model of dark energy, the
 curvature constraint can be improved. However, the precision depends on
 which dark energy model we adopt. If we assume the dark energy model
 parametrized by its equation of state, $w_{\rm DE}(z)=w_0+w_a(1-a)$
 around the fiducial model with the cosmological constant, the curvature
 constraint can be significantly improved. If we restrict ourselves to
 the cosmological constant ($w_{\rm DE}=-1$), the best curvature
 constraint is achieved (as denoted by the lower curve), but the
 constraint is model dependent.}  \label{fig:err_ok}
\end{figure*}
\begin{figure*}
    \centering
    \includegraphics[width=5in]{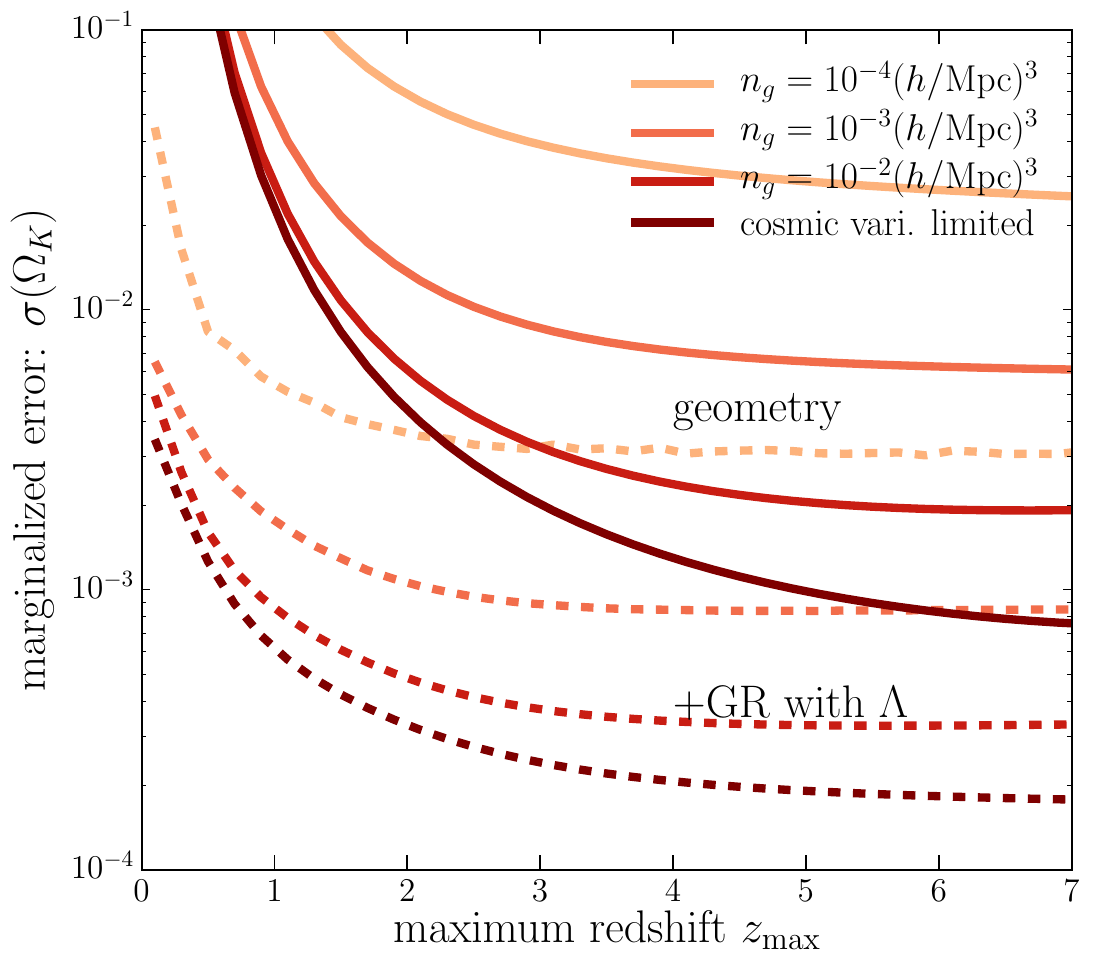}
 \caption{Similar to the previous figure, but this figure shows how the
 curvature constraints are degraded for a galaxy survey with a finite
 number density of galaxies in each redshift bin. Here we consider the
 same number density in different redshift bins for simplicity.}
 \label{fig:err_ok_ng}
\end{figure*}

\subsection{Survey parameters}

Here we assume an ideal survey to estimate the fundamental limit of the
curvature estimation via the BAO measurements. That is, we assume a
galaxy survey with full-sky coverage and a sufficiently high number
density of sampled galaxies in each redshift bin up to a given maximum
redshift $z_{\rm max}$, without any gap. We will study how the curvature
determination is degraded when considering a survey with a finite number
density of galaxies.

\subsection{Forecasts}

Figure~\ref{fig:dist} shows the forecasts of the distance measurements
with a cosmic-variance-limited survey up to $z_{\rm max}=7$. The shaded
region around each redshift bin with width $\Delta z=0.2$ denotes the $\pm
1\sigma$ uncertainty in the angular diameter distance including
marginalization (Eq.~\ref{eq:covDD}). The accuracy of the Hubble
distance is only slightly worse than the error in $D_A$ (at the level of
$10\%$).
As described in Sec.~\ref{sec:BAO}, we include both the BAO
feature and the broadband feature of the galaxy power spectrum (the AP test)
for the distance forecasts. We confirmed that the distance accuracies
are mainly from the BAO features. We should also stress that, once a
sufficient number of redshift slices, up to $z\sim 1$ for our setting,
are included, the BAO information of the galaxy survey allows a more
accurate determination of the sound horizon scale than the Planck prior,
which explains why the distance accuracies get saturated at
$z\simgt 1$.

Using the forecasts of the distance measurements, we show in
Fig.~\ref{fig:err_ok} the expected accuracy of the curvature estimation,
$\sigma(\Omega_K)$.  Here we consider a cosmic-variance-limited galaxy
survey over a range of redshift $z=[0,z_{\rm max}]$ and show the
forecasts as a function of the maximum redshift $z_{\rm max}$.  Note
that the error $\sigma(\Omega_K)$ is free of the uncertainty in the
Hubble parameter $h$, because the uncertainty in $h$ is absorbed when
estimating the curvature parameter from a combination of the radial and
angular distances. The top curve shows the fundamental limit, estimated
from Eq.~(\ref{eq:covKK}), in the sense that the constraint is purely
based on the BAO distance measurements and is less sensitive to the theory
of structure formation (other than the BAO scale) such as the galaxy bias
and is free of any model of dark energy. The accuracy can be better than
$0.1\%$, if a cosmic-variance-limited survey up to $z_{\rm max}\simgt 5$
is available. Since the curvature has a relatively larger impact on the
expansion history up to higher redshifts than the cosmological
constant, adding the BAO distance measurements at higher redshift keeps
improving the curvature constraint.

If we employ Einstein gravity (GR), which relates
the structure of spacetime to the energy-matter content of the Universe
through the Einstein equations, the curvature constraint can be
improved, although the constraint is now model-dependent. At redshifts
relevant for galaxy surveys, the Hubble expansion rate is given as
\begin{equation}
 H(z)^2=H_0^2\left[
\Omega_{\rm m0}(1+z)^3 + \Omega_K(1+z)^2+\Omega_{\rm DE}(z)
	     \right].
\end{equation}
Then the Hubble and angular-diameter distances are specified by the
density parameters (see Eqs.~\ref{eq:dc} and \ref{eq:da_dc}). However,
the nature of dark energy is another mystery of the universe. For
comparison, we here employ a phenomenological model of dark energy that is
parametrized in terms of its equation-of-state parameters as
\begin{equation}
 w_{\rm DE}(z)=w_0+w_a(1-a),
\end{equation}
where $w_0$ and $w_a$ are parameters and constant in time. In this case,
the redshift evolution of dark energy is given as $\rho_{\rm
De}(z)\propto a^{-3(1+w_0+w_a)}e^{-3w_a(1-a)}$ with $1+z=1/a$.  We
should emphasize that this model acts as a ``strong'' prior, restricting
the degrees of freedom of the dark energy model to a specific model. The
redshift dependence of the above dark energy model, around the fiducial
cosmological constant model, is different from that of the curvature
($a^{-2}$), the BAO distance measurements determine the curvature and
the dark energy parameters simultaneously. If the dark energy has an
(even very small) additive contribution, such as $\delta \rho_{\rm
DE}\propto 1/a^2$, the contribution leaves a strong degeneracy with the
curvature.

Nevertheless it would be useful to estimate the accuracy of the curvature
determination from the cosmic-variance-limited BAO measurements when
assuming the above dark energy model. To do this, we use the method in
Ref.~\cite{SeoEisenstein:03} to propagate the accuracies of BAO distance
measurements into estimations of cosmological parameters. To be more
precise, we compute the submatrix of the inverted Fisher matrix of CMB
plus the BAO measurements:
\begin{eqnarray}
&&   \bm{F} \overset{\mbox{invert}}{\Longrightarrow}
 [\bm{F}]^{-1}
 \overset{\mbox{use submatrix}}{\Longrightarrow}
  [\bm{F}]^{-1}_{M'N'}.
\end{eqnarray}
The submatrix $F_{M'N'}$ includes marginalization over other parameters
such as galaxy bias (see Eq.~\ref{eq:parameters}). Then the elements
denoted by $M',N'$ include parameters that determine the distances among
the set of parameters (Eq.~\ref{eq:parameters}), and then we make the
parameter forecasts as follows:
\begin{eqnarray}
&& p_{M'}=\{\Omega_{\rm m0},\Omega_{\rm m0}h^2, D_A(z_i), D_H(z_i) \}\nonumber\\
 && \overset{\mbox{ }}{\Longrightarrow}
F_{{\rm DE},ij} \equiv  \frac{\partial p_{M'}}{\partial q_i}
  F_{M'N'}  \frac{\partial p_{M'}}{\partial q_j}\nonumber\\
 && \overset{\mbox{ }}{\Longrightarrow}
  q_i=\{\Omega_{\rm m0}h^2, \Omega_{\rm DE0}, \Omega_{K}, w_0, w_a\}.
\end{eqnarray}
The second line denotes a projection of the Fisher matrix onto a new
parameter space denoted by $q_i$.
Then the third line denotes a
derivation of the marginalized error on the parameter including the
curvature and the dark energy parameters. 

The dashed curves in Fig.~\ref{fig:err_ok} show the marginalized error of
the curvature parameter when assuming GR to model the Hubble expansion
history and employing the dark energy model, around the fiducial
cosmological constant model. By restricting the analysis to a narrower
range of dark energy, the curvature constraint can be improved. If
assuming the cosmological constant model (the  $\Lambda$CDM model), the
cosmic-variance-limited accuracy can be as high as
$\sigma(\Omega_K)\simeq 2\times 10^{-4}$. The figure shows that the
curvature accuracy gets saturated more quickly at $z\simgt 1$, compared
to the solid curve without any prior on dark energy. This is for
two reasons. First, if the dark energy model is employed, a degeneracy
between the curvature and the dark energy parameters can be well broken
by adding the BAO measurements in multiple redshifts, where the dark
energy around the cosmological constant affects the expansion history
only up to $z\sim 1$. Second, the sound horizon scale can be well
determined by adding the BAO measurements up to $z\sim 1$ as we stated
above. For these reasons, the BAO measurements at $z\simgt 1$ add
relatively less information on the curvature when the dark energy models
are employed, 
compared to the case
without any prior on dark energy (solid curve).

In reality, the accuracy of the curvature determination is degraded by a
finite number of sample galaxies for a spectroscopic
survey. Figure~\ref{fig:err_ok_ng} shows how the curvature estimation is
degraded as a function of the number density of galaxies. Here we assume
a constant number density of galaxies in each redshift bin for
simplicity. The planned galaxy surveys (such as the Subaru PFS, Euclid
and WFIRST) are designed so as to satisfy a requirement of
$\bar{n}P_g(k)\simgt 1$ at $k\simeq 0.1$--0.2$h/{\rm Mpc}$,
corresponding to a number density of $\bar{n}_g\simeq 10^{-3}~(h/{\rm
Mpc})^3$ \citep{Takadaetal:14,WFIRST:15}. In this case,
$\sigma(\Omega_K)\simeq 0.006$ or 0.0009 for the pure geometry and for
the GR+$\Lambda$ case, respectively. However, note that these future
surveys aim at using mainly emission-line galaxies such as [OII] and/or
H$\alpha$ emitters as tracers of large-scale structure. The
emission-line galaxies are a tiny fraction of imaging galaxies at a
depth of $i\sim 25$mag (less than $1\%$). By having a wider wavelength
coverage (e.g., up to infrared wavelengths as proposed by the SPHEREx
mission \cite{SPHEREx}) and/or improving the sensitivity of the spectrograph,
a high number density of $\sim 10^{-2} (h/{\rm Mpc})^2$ for a redshift
galaxy survey will be feasible in principle. Intergalactic medium surveys,
such as Lyman-$\alpha$ forests \citep[e.g., Ref.][]{Slosaretal:13} or radio
intensity mapping \citep[e.g., Ref.][]{Bulletal:15} can also be combined with
optical and infrared redshift surveys. Hence a (nearly)
cosmic-variance-limited
 BAO survey with $\bar{n}\sim 10^{-2}(h/{\rm Mpc})^3$ is not
impossible, although challenging.

\section{Discussion}
\label{sec:discussion}

We have estimated the accuracy of the geometrical determination of the
spatial curvature from the BAO distance measurements. The BAO is unique
in the sense that it allows for a measurement of the radial distance at the
redshift of the galaxy survey, while other methods -- such as type-Ia supernova
and gravitational lensing -- measure the luminosity distance which is
equivalent to the angular diameter distance. Hence, combining the radial
and angular diameter distances allows us to constrain the spatial
curvature without making many assumption about the theory of structure
formation (other than the BAO scale) such as galaxy bias or any model of
dark energy. We showed that an all-sky, cosmic-variance-limited galaxy
survey covering up to a high redshift of $z\simgt 4$ allows for a
curvature determination to an accuracy of $\sigma(\Omega_K)\simeq
10^{-3}$. If we assume a simple model of dark energy that is
parametrized by two equation-of-state parameters, $w(a)=w_0+w_a(1-a)$,
it allows for an accuracy of $\sigma(\Omega_K)\simeq \mbox{a few}\times
10^{-4}$, although the constraint is considered model dependent in this
case.

If there is another method of measuring the Hubble expansion rate at
high redshift, it might further enable us to improve the curvature
constraint based on the method we discussed in this paper. For example,
if ``red-envelope'' galaxies -- which have formed their stellar population
at higher redshift at $z\simgt 2$ -- can be used as a cosmic chronometer
(as demonstrated in Ref.~\cite{Sternetal:10}), the ages of such red
galaxies from detailed spectroscopic observations can be used to
estimate the Hubble expansion rate at the galaxy redshift. However, this
method still seems to be limited by astrophysical systematic effects; a
further careful study needs to be done. The spectroscopic observations of
an extremely large-aperture telescope, in combination with the optical
frequency comb technique, in principle allow for a precision measurement of
the Hubble expansion rate in the high-redshift Universe
\citep[e.g., Ref.][]{Liskeetal:08}.

There is an interesting, physical contamination to the estimation of the
curvature. A coherent density contrast across a survey region or a
local patch in an inhomogeneous universe can appear as a curved universe
even if the global geometry is flat \cite{Baldaufetal:11,Daietal:15}. As
can be found from Eq.~(43) in Ref.~\cite{Lietal:14a}, the induced,
apparent curvature is expressed as $\Omega_K^{\rm local}\sim \delta_{\rm
b}$ with an $O(1)$ prefactor, where $\delta_{\rm b}$ is the coherent
density contrast. If a local universe is embedded into a coherent
underdensity region up to $z\sim 1$, the comoving survey volume is
estimated as $V(<z=1)\sim 50~ ({\rm Gpc}/h)^3$ for an all-sky survey.
The $\Lambda$CDM model predicts $\sigma_{\rm b}\sim \mbox{a few}\times
10^{-4}$ for the expected rms of the density contrast for the volume, as
can be found from Fig.~1 in Ref.~\cite{TakadaHu:13} \citep[also
see Ref.][]{Lietal:14b}. Hence such a $1\sigma_{\rm b}$ negative density
contrast could degrade a determination of the global curvature or
completely mimic the global curvature, if the true curvature is as
small as $\Omega_K\sim 10^{-4}$. However, as stressed in
Refs.~\cite{TakadaHu:13,Lietal:14b,TakadaSpergel:13,Schaanetal:14} such
a coherent density contrast also causes characteristic, scale-dependent
modifications in the growth of all small-scale structures (such as weak
lensing and cluster abundance) compared to their flat $\Lambda$CDM
expectations. Hence, although the apparent curvature effects themselves
are very intriguing to explore, the effects can in principle be
distinguished by combining the geometrical BAO measurements with probes
of large-scale structure growth. In addition, even if such a coherent
underdensity contrast exists, it would be very difficult (or there would
only be an
incredibly small chance) to have a situation that we on the Earth are
located at the center of such a large-scale void. Hence having a
wider-area coverage to explore anisotropic curvature effects on the sky
as well as having a larger redshift leverage to explore the void
boundary should help distinguish the apparent curvature effects
observationally.

Constraining the curvature together with properties of the primordial
scalar perturbations and the primordial gravitational wave is an
important direction to explore. If a nonzero spatial curvature is found
from cosmological observations, it will definitely change our view and
understanding of the Universe, so it is worth exploring with any
possible means of current or future observational data sets.

 \begin{acknowledgements}
  We thank Gary Bernstein, Neal Dalal, Chris Hirata, Eiichiro Komatsu,
  Hitoshi Murayama, Yasunori Nomura, Hirosi Ooguri, and Misao Sasaki for
  useful discussion.  MT is supported by the World Premier International
  Research Center Initiative (WPI Initiative), MEXT, Japan, by the FIRST
  program “Subaru Measurements of Images and Redshifts (SuMIRe)”,
  CSTP, Japan, by a MEXT Grant-in-Aid for Scientific Research on
  Innovative Areas, ``Why does the Universe accelerate? - Exhaustive
  study and challenge for the future -'' (No.~15H05893 and 15K21733), by
  a Grant-in-Aid
  for Scientific Research from the JSPS Promotion of Science
  (No. 23340061 and 26610058), and by the JSPS Program for Advancing
  Strategic International Networks to Accelerate the Circulation of
  Talented Researchers. Part of the research described in this paper was
  carried out at the Jet Propulsion Laboratory, California Institute of
  Technology, under a contract with the National Aeronautics and Space
  Administration. OD thanks IPMU for its generous hospitality. This work
  is also supported in part by the National Science Foundation under
  Grant No. PHYS-1066293 and the hospitality of the Aspen Center for
  Physics.
 \end{acknowledgements}
\bibliography{ms}

\end{document}